\documentclass[%
 reprint,
 amsmath,amssymb,
 aps,prl
]{revtex4-2}

\usepackage{graphicx}
\usepackage{dcolumn}
\usepackage{bm}

\renewcommand\apj{{ApJ}}
\newcommand\apjl{{ApJL}}     
\newcommand\apjs{{ApJS}}
\newcommand\aap{{A\&A}}
\renewcommand\nat{{Nature}}
\newcommand\physrep{{PhR}}
\def\ie{{\em i.e., }}
\def\be{\begin{equation}}
\def\bea{\begin{eqnarray}}
\def\eea{\end{eqnarray}}
\def\ee{\end{equation}}
\begin{document}

\preprint{APS/123-QED}

\title{Photons' scattering in a relativistic plasma with velocity shear: generation of high energy power-law spectra}

\author{Mukesh K. Vyas}
 \email{mukeshkvys@gmail.com}
\author{Asaf Pe'er}%
\affiliation{%
 Department of Physics, \\Bar Ilan University, \\ 
Ramat Gan, \\
Israel, 5290002 \\
}%


%

\date{\today}

\begin{abstract}
A high energy power law is a common feature in the spectra of many astrophysical objects. We show that the photons in a relativistic plasma with a variable Lorentz factor go through repeated scattering with electrons to gain energy. The escaped population of photons naturally produces a power-law-shaped spectrum making it an anisotropic analogue to the conventional Fermi acceleration mechanism of charged particles. Thus, this mechanism provides a natural alternative to current explanations of high energy power-law spectra via synchrotron or thermal Comptonization.  The model is applicable to any relativistic plasma beam with an arbitrary Lorentz factor profile. We implement the theory to GRB prompt phase and show that the obtained range of the photon indices is compatible with the observed values. Therefore, the observed high energy spectral indices provide a unique indicator of the jet structure.
\end{abstract}

\maketitle

A power-law spectrum at high energies appears ubiquitously in several astrophysical sources like active galactic nuclei [AGNs] \cite{Nandra&Pounds1994,Reeves&Turner2000,Page.etal.2005}, and gamma-ray bursts [GRBs] \cite{Band.etal.1993,Kaneko.etal.2006ApJS..166..298K,Bosmjak.etal.2014A&A...561A..25B,pe'er2015AdAst2015E..22P,Preece.etal.1998ApJ...496..849P,Preece.etal.2000ApJS..126...19P,Barraud.etal.2003A&A...400.1021B}. These objects are characterized by trans- or highly relativistic jets, namely having Lorentz factor $\Gamma \gg 1$. Such jets are produced by collapse of a massive star, merger of two compact objects (in the case of GRBs)\citep{Levinson&Eichler1993ApJ...418..386L,Woosley1993AAS...182.5505W,MacFadyen&Woosley1999ApJ...524..262M,2022Univ....8..294V}, or through an accretion disc surrounding a black hole \cite{Junor1999Natur.401..891J,Doeleman.etal.2012Sci...338..355D,2015MNRAS.453.2992V,2017MNRAS.469.3270V,2018A&A...614A..51V,Truong&Newman2018MNRAS.477.1803L,2018JApA...39...12V,2019MNRAS.482.4203V,Fukue2021MNRAS.503.1367F,Aneesha&Mandal.2020A&A...637A..47A}. Although the geometric shape of these jets is uncertain, numerical modeling shows some typical jet profiles where the jet's Lorentz factor is a function of its polar angle, [\ie $\Gamma=\Gamma(\theta)$] which may be universal \cite{Zhang.etal.2003ApJ...586..356Z,Lundman.etal.2013MNRAS.428.2430L,Pe'er&Ryde2017IJMPD..2630018P}. This internal jet structure with a velocity shear within the plasma implies that the photons emitted deep inside the flow are scattered in regions with different velocities before escaping. As we show here, a part of the photon population gains energy, resulting in a power-law-shaped spectrum at high energies. This process draws similarity with second-order Fermi acceleration of charged particles interacting with randomly moving magnetic irregularities \cite{Blandford&Eichler1987PhR...154....1B}. 
In this process, the photons gain energy from the bulk kinetic energy of the jet itself. The origin of this effect is independent of the electron temperature $T_{\rm el}$, and it will operate as well with $T_{\rm el} = 0$. This mechanism, thus, is a viable alternative to the known mechanisms that are invoked in producing the power-law spectra, such as synchrotron \cite{Zdziarski.etal.2017MNRAS.471.3657Z,Burgess.etal.2014ApJ...784...17B,Yu.etal.2015A&A...573A..81Y,pe'er2015AdAst2015E..22P,Tavani1996ApJ...466..768T,Cohen.etal.1997ApJ...488..330C,Schaefer.etal.1998ApJ...492..696S,Frontera.etal.2000ApJS..127...59F,Wang.etal.2009ApJ...698L..98W,Ryde&Pe'er.2009ApJ...702.1211R} or inverse Compton from power-law accelerated electrons, or by thermal Comptonization \cite{Vyas.etal.2021ApJ...908....9V, vyas.etal.2021.predictingAPJL}.   Within this framework, the dependence of the emerging high-energy spectral indices on the dynamical variables of the system thus provides a novel signature of the jet structural profile, $\Gamma=\Gamma(\theta)$.

An astrophysical jet can be pictured as a relativistic plasma beam expanding in space at angles $\theta \leq \theta_{\rm out}$ with the propagation axis being at $\theta=0$. Assuming cylindrical symmetry, we consider a Lorentz factor profile of the plasma to be a function of the radial and angular coordinates, $\Gamma=\Gamma(r,\theta)$.  

\begin {figure}[h]
\begin{center}
 \includegraphics[width=6cm, angle=0]{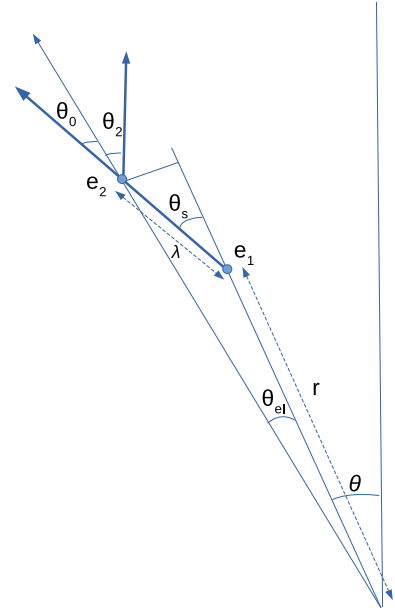}
\caption{Geometry of scattering in lab frame in spherical coordinates $r,\theta,\phi$. The origin lies at the centre of the star. $\theta$ is the location of a certain scattering event of the photon with electron $e_1$ at $r$ radial distance. $\theta+\theta_{\rm el}$ is the angular location of the next scattering event with electron $e_2$. $\theta_{\rm s}$ and $\theta_2$ are the scattering angles in the lab frame for both scattering events, respectively.}
\label{lab_geom_2}
 \end{center}
\end{figure}

The conditions deep inside the jet, namely high density, high temperature and strong magnetic fields \cite{pe'er2015AdAst2015E..22P} imply a bright emission of photons at some characteristic inner radii $r_0$. These photons, then propagate and interact with the electrons inside the plasma until they escape. In describing the scattering process, we omit the random thermal motion of the electrons, an assumption which is valid as long as adiabatic losses imply that the Lorentz factor associated with the thermal motion is small with respect to the bulk Lorentz factor (\ie $\gamma << \Gamma$).

As the emitted photons propagate inside the expanding jetted plasma, they gain energy (on average) by multiple scattering with the electrons.  Consider a photon that undergoes a scattering with an electron having a (bulk) Lorentz factor $\Gamma$ at location $(r,\theta)$, and emerges with energy $\varepsilon$ at an angle $\theta_{\rm s}$ with respect to the direction of the electron (all quantities are measured in the lab frame; see Figure  \ref{lab_geom_2}).  

After being scattered, the photon interacts with a second electron having Lorentz factor $\Gamma_2$ at a new location, and is scattered at an angle $\theta_{2}$ with emerging energy $\varepsilon_2$. 
We consider the Thomson limit, namely that the energy of the photon before and after the second scattering, as measured in the (second) electron's rest frame, $\varepsilon'[= \Gamma_2 \varepsilon (1-v_2 \cos \theta_{0})$] and $\varepsilon_2'[ =  \Gamma_2 \varepsilon_2 (1-v_2 \cos \theta_{2})$] are equal, \ie $\varepsilon_2'=\varepsilon'$.
Here $v_2$ is the velocity of the second electron (in units of light speed) and  $\theta_0$ is the angle between the electron's direction and the incoming photon's direction in the lab frame. Denoted by $\theta_{\rm el}$ is the angular shift of the photon location between the two scattering events in the lab frame. From the geometry, one has $\theta_0=\theta_{\rm s}-\theta_{\rm el}$, or $\varepsilon'= \Gamma_2\varepsilon [1-v_2 \cos (\theta_{\rm s}-\theta_{\rm el})]$.
The energy gain in this scattering process is therefore,

\be 
g=\frac{\varepsilon_2}{\varepsilon}=\frac{1-v_2 \cos (\theta_{\rm s}-\theta_{\rm el})}{1-v_2 \cos (\theta_{2})}.
\label{eq_gain}
\ee

As the jets are relativistic, the average scattering angles are $\theta_{\rm s}\approx 1/\Gamma$ and $\theta_2 \approx 1/\Gamma_2$. Further, for a narrow plasma beam, $\cos \theta_0 \approx 1-\theta_0^2/2$ and $\cos \theta_2 \approx 1-\theta_2^2/2$, \ie 
\be 
g=\frac{1}{2}\left[1+\Gamma_2^2 (\theta_{\rm s}-\theta_{\rm el})^2\right].
\label{eq_gain_2}
\ee
%
If the local mean free path measured in the lab frame is $\lambda(r,\theta)$, then
\be 
\tan \theta_{\rm el} = \frac{\lambda \sin \theta_{\rm s}}{r+\lambda \cos \theta_{\rm s}} = \frac{a \theta_{\rm s}}{1+a \cos \theta_{\rm s}}\Rightarrow\frac{\theta_{\rm el}}{\theta_{\rm s}}\approx \frac{a}{1+a},
\label{eq_te}
\ee
where $a \equiv \lambda/r$.
Using it in equation \ref{eq_gain_2}, the energy gain becomes
\begin{equation}
g\approx\frac{1}{2}\left[1+\left(\frac{\Gamma_2}{\Gamma}\right)^2 \frac{1}{\left(1+a\right)^2}\right].
\label{eq_gain_4}
\end{equation}
For a relatively small mean free path in an optically thick and dense plasma, $\lambda \ll r$, one has $\Gamma_2=\Gamma+\delta \Gamma$,
namely
\be 
\frac{\Gamma_2}{\Gamma}=1+\frac{\delta \Gamma}{\Gamma}=1+\sum \frac{\partial \log \Gamma}{\partial x^i }\delta x^i.
\ee
Here $x^i$ are spatial coordinates. Using this expression in Equation \ref{eq_gain_4}, the gain becomes
\be 
g(r,\theta) \approx\frac{1}{2}\left[1+\left[1+\sum \frac{\partial \log \Gamma}{\partial x^i }\delta x^i\right]^2 \frac{1}{\left(1+a\right)^2}\right].
\label{eq:6}
\ee
This expression implies that the photon can gain ($g>1$) or lose energy ($g<1$) in a single scattering event. A sharp and positive fractional Lorentz factor gradient ($\partial \log \Gamma / \partial x^i$) may be sufficiently steep to overcome the adiabatic losses due to the plasma expansion [the $(1+a)^{-2}$ term]. It results in a net energy gain for the photon (\ie $g>1$). On the other hand, the net gain due to the shear may be insufficient, in which case the energy loss due to the expansion would dominate, leading to a net energy loss, (\ie $g<1$).

The expression in Equation \ref{eq:6} can be simplified by considering azimuthal symmetry, for which $\partial \Gamma/\partial \phi = 0$. Furthermore, for scattering at an average angle $1/\Gamma$ in the lab frame, $\delta r$ is always positive, while the sign of $\delta \theta (\partial \Gamma/\partial \theta)$ depends upon the sign of $\delta \theta$. A photon can scatter away from the propagation axis ($\theta=0$) with an average angle $\delta \theta$ = $\theta_{\rm el}$, or it can scatter towards the axis with an average angle $\delta \theta$ = $-\theta_{\rm el}$. Both cases have equal probabilities. If we denote the gain in these two cases by $g_+$ and $g_-$, the average gain in a single scattering at location ($r$, $\theta$) is

$$ 
g_{\rm a}(r,\theta)= \frac{g_++g_-}{2}.
$$

The average energy gain depends both on the Lorentz factor gradient and the radial location via the parameter $a$.
The value of $a$ is calculated by noting that the mean free path $\lambda$ transforms from the comoving frame ($\lambda_0$) to the lab frame as
\be 
\lambda = \frac{\lambda_0}{\Gamma [1-v \cos (\theta_{\rm s})]} =\frac{1}{\Gamma n_{\rm e}'\sigma_T [1-v \cos (\theta_{\rm s})]}\approx\frac{\Gamma}{n_{\rm e}'\sigma_T}.
\label{eq_mfp}
\ee
Here $\sigma_T$ is the Thomson scattering cross-section, $n_{\rm e}'$ is electron number density in the local comoving frame of the first electron, and $v$ is the bulk speed of the first electron in terms of light speed. Hence, $a={\Gamma}/{rn_{\rm e}'\sigma_T}$.

The expectation value of the photon energy gain in the plasma is evaluated by integrating the average gain over the entire region of scattering,
\be 
\bar{g} = \frac{1}{V}\int_{V}g_{\rm a}(\theta,r)  dV.
\label{eq_int_gain}
\ee
Here, $V$ is the volume of total region within angular boundaries $0 ... \theta_{\rm out}$ and radial extend $r_0 \leq r \leq r_{\rm ph}$, where $r_{\rm ph}$ is the photospheric radius at which the photons escape to infinity. This radius depends upon the specific jet geometry.

The average probability of the photon to have a scattering without escape within the jet is the probability $P(r,\theta)$ averaged over the available volume $V$ of scattering where velocity shear is present, \ie
\be
\bar{P} = \frac{1}{V}\int_{V}  P(r,\theta) dV.
\label{eq_average_P_final}
\ee  
A photon can escape the scattering region through the photospheric radius or from the angular jet boundary at $\theta=\theta_{\rm b}$ (In this case, it is $\theta_{\rm out}$). In the former case, the escape probability of the photon is $\exp(-\tau_1)=\exp(-r_{\rm ph}/r)$, where $\tau_1$ is the photospheric optical depth. In the latter case, the probability of escape through the angular boundary ($\theta_{\rm b}$) is calculated by estimating the optical depth along direction $\theta_{\rm s}$ (see Figure \ref{lab_geom_2}),
\be 
\tau_2=\int_0^{s_0}\Gamma(1-v \cos \theta_{\rm s})n_{\rm e}' \sigma ds.
\label{eq_tau_2}
\ee
Here, $ds$ is a length element along $\theta_{\rm s}$, and the integration boundary is $s_0=\left(|\theta-\theta_{\rm b}|\right)r/\theta_{\rm s}$. 
The photon entertains the minimum of these two optical depths $\tau_1$ and $\tau_2$ to escape the scattering region, \ie $\tau=\min(\tau_1,\tau_2)$. Thus, the probability that the photon would escape is $P_{\rm e}(r,\theta) = \exp[-\tau(r,\theta)]$, while the probability for it to have next scattering inside the beam is $P(r,\theta)=1-{P_{e}}$. Using it in Equation \ref{eq_average_P_final} enables us to estimate $\bar{P}$.

To calculate the spectrum of the escaped photons, consider that $N_0$ photons are homogeneously injected at $r=r_0$ \footnote{This is a valid approximation, as $r_0<<r_{\rm ph}$.}.
After scattering $k$ times, $N = N_0\bar{P}^k$ photons are left within the scattering region. 
After $k^{th}$ scattering, the photon's average energy is $\varepsilon_{\rm k}=\varepsilon_{\rm i} \bar{g}^k $, implying 
\be 
\frac{N}{N_0} = \left(\frac{\varepsilon_{\rm k}}{\varepsilon_{\rm i}}\right)^{\beta'},
\label{eq_int_spectrum}
\ee
where $\beta'= \frac{\ln \bar{P}}{\ln \bar{g}}$. 
The resulting photon index of the escaped photons is 
\be 
\beta=\beta'-1 = \frac{\ln \bar{P}}{\ln \bar{g}}-1.
\label{eq_photon_ind}
\ee
Thus, for a given set of dynamic parameters, the high energy part of escaped photons' spectrum is characterized by a power-law with photon index $\beta$. 
For a defined Lorentz factor profile and angular boundaries ($\theta_i$, $\theta_{\rm out}$), radial boundaries $r_0({r,\theta,\phi})$ and $r_{\rm ph}({r,\theta,\phi})$, the photon index of the observed spectra can be evaluated directly from the equations above. The treatment is valid as long as the plasma is relativistic (So that Equation \ref{eq_gain_2} holds). 
\begin {figure}[h]
\begin{center}
 \includegraphics[width=8cm, angle=0]{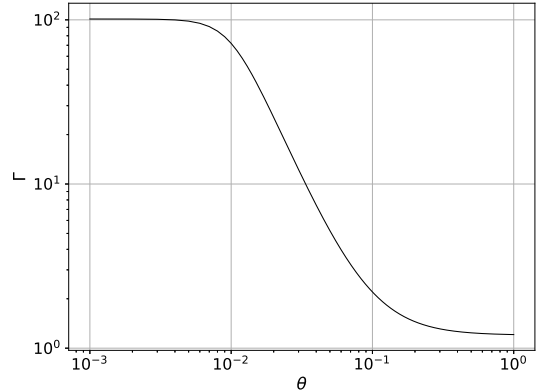}
\caption{Lorentz factor ($\Gamma$) profile of the jet characterized by equation \ref{eq_gamma_1} with parameters $p=2.0$, $\theta_{\rm j}=0.01$ rad, $\Gamma_0=100$ and $\Gamma_{\rm min}=1.2$. $\theta_{\rm e}=\theta_{\rm j}\Gamma_0^{1/p}$. The inner jet region is for $\theta<\theta_{\rm j}$ while outer region extends beyond $\theta_{\rm e}$. The region bounded within $\theta_{\rm j}-\theta_{\rm e}$ harbours an effective velocity shear leading to photon energy gain.}
\label{lab_jet_profile}
 \end{center}
\end{figure}

As a test case, we apply the model to the prompt phase of GRBs where, within the framework of the ``collapsar" model \cite{Levinson&Eichler1993ApJ...418..386L,Woosley1993AAS...182.5505W,MacFadyen&Woosley1999ApJ...524..262M},  a jet is launched from the centre of a collapsing star.  Once erupted, the jet propagates along direction $\theta=0$ above the stellar surface. As shown by Zhang et al \cite{Zhang.etal.2003ApJ...586..356Z}, a relativistic GRB jet harbours an angular structure where the dependency of their Lorentz factor ($\Gamma$) on polar angle $\theta$ can be approximated by \cite{Lundman.etal.2013MNRAS.428.2430L,Pe'er&Ryde2017IJMPD..2630018P} 
\be 
\Gamma(\theta)= \Gamma_{\rm min}+\frac{\Gamma_{0}}{\sqrt{\left(\frac{\theta}{\theta_{\rm j}}\right)^{2p}+1}}.
\label{eq_gamma_1}
\ee 
This profile is plotted in Figure \ref{lab_jet_profile}.
Here, $\Gamma_{\rm min}$ and $\Gamma_0$ are constants with $\Gamma_0>\Gamma_{\rm min}$, and $p$ is the jet profile index that shows the sharpness of the velocity gradient between shear layers of the jet. The inner region of the jet ($\theta<\theta_{\rm j}$) approaches a maximum Lorentz factor $\Gamma_0$ while the region outside $\theta_{\rm e} \sim \theta_{\rm j}\Gamma_0^{1/p}$ asymptotically reaches $\Gamma_{\rm min}$. 

For such a jet, the comoving electron number density is $n_{\rm e}'=L/4\pi m_{\rm p} c^3 v \Gamma^2 r^2$,
where $L$ is the jet luminosity and $v$ is the bulk jet speed in units of $c$. One therefore concludes that $a={v r}/{2r_{\rm ph}}$ and the photospheric radius for an on axis observer 
$r_{\rm ph} ({\rm observer's~ angle~} \theta_{\rm o}=0)$ is given by $ r_{\rm ph} = {\sigma L}/{8\pi c^3 m_{\rm p} \Gamma^3}$  [see Lundman et al. \cite{Lundman.etal.2013MNRAS.428.2430L} for further details].
The optical depth for photon propagation along the radial direction is a strong function of the propagation angle, $\theta$. In the inner jet region ($\theta<\theta_{\rm j}$), the optical depth along the photon's path is significantly smaller than the outer region [as $\tau_1\propto1/\Gamma^3$]. Thus, the photons easily escape the jet once they reach the inner region.
Hence, the angular limit $\theta_{\rm b}=\theta_{\rm j}$ is an effective boundary for photon escape for the case of a GRB jet. 
\begin{figure*}
\begin{center}
  \includegraphics[width=8cm, angle=0]{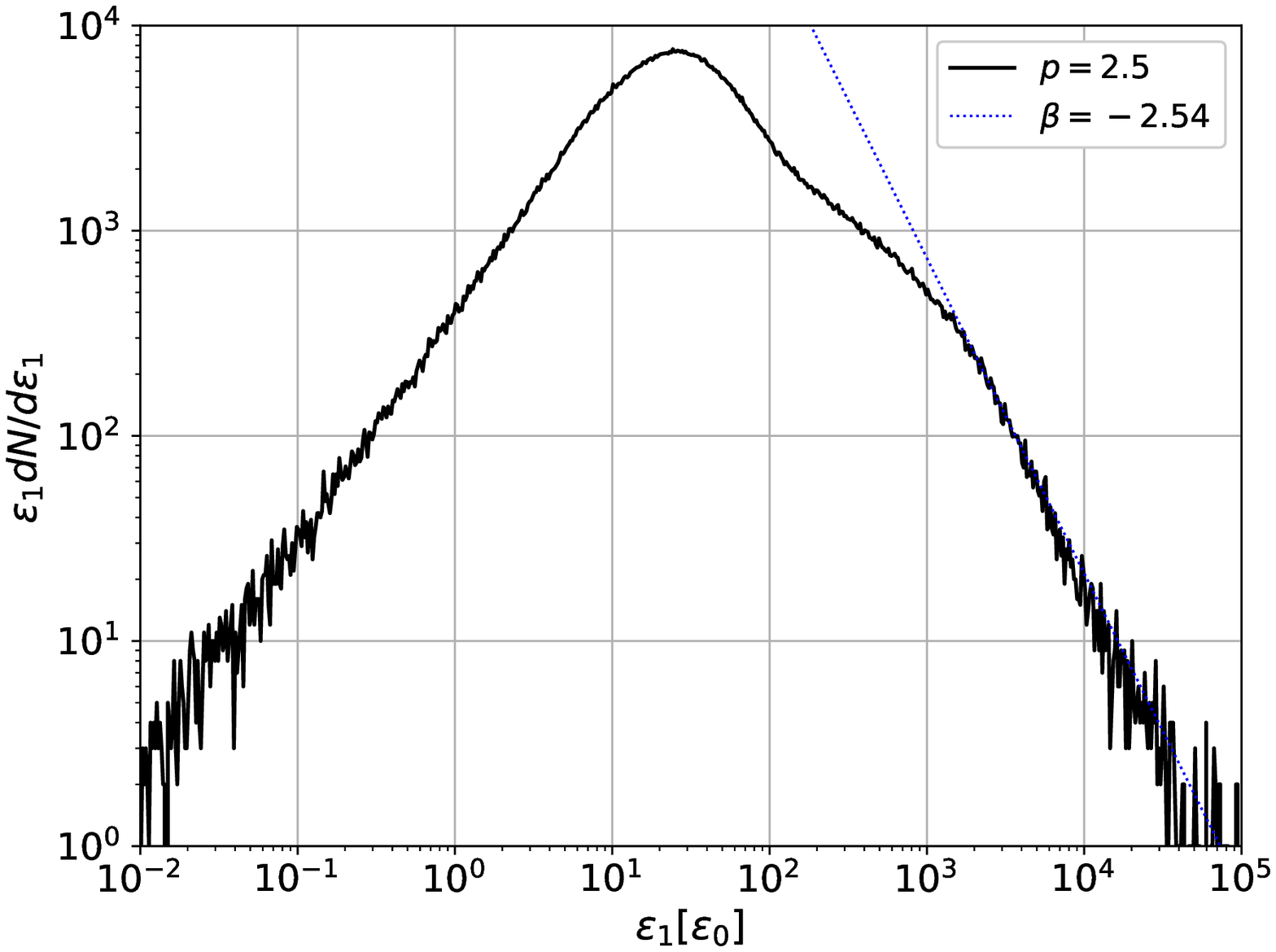}
            \includegraphics[width=8cm, angle=0]{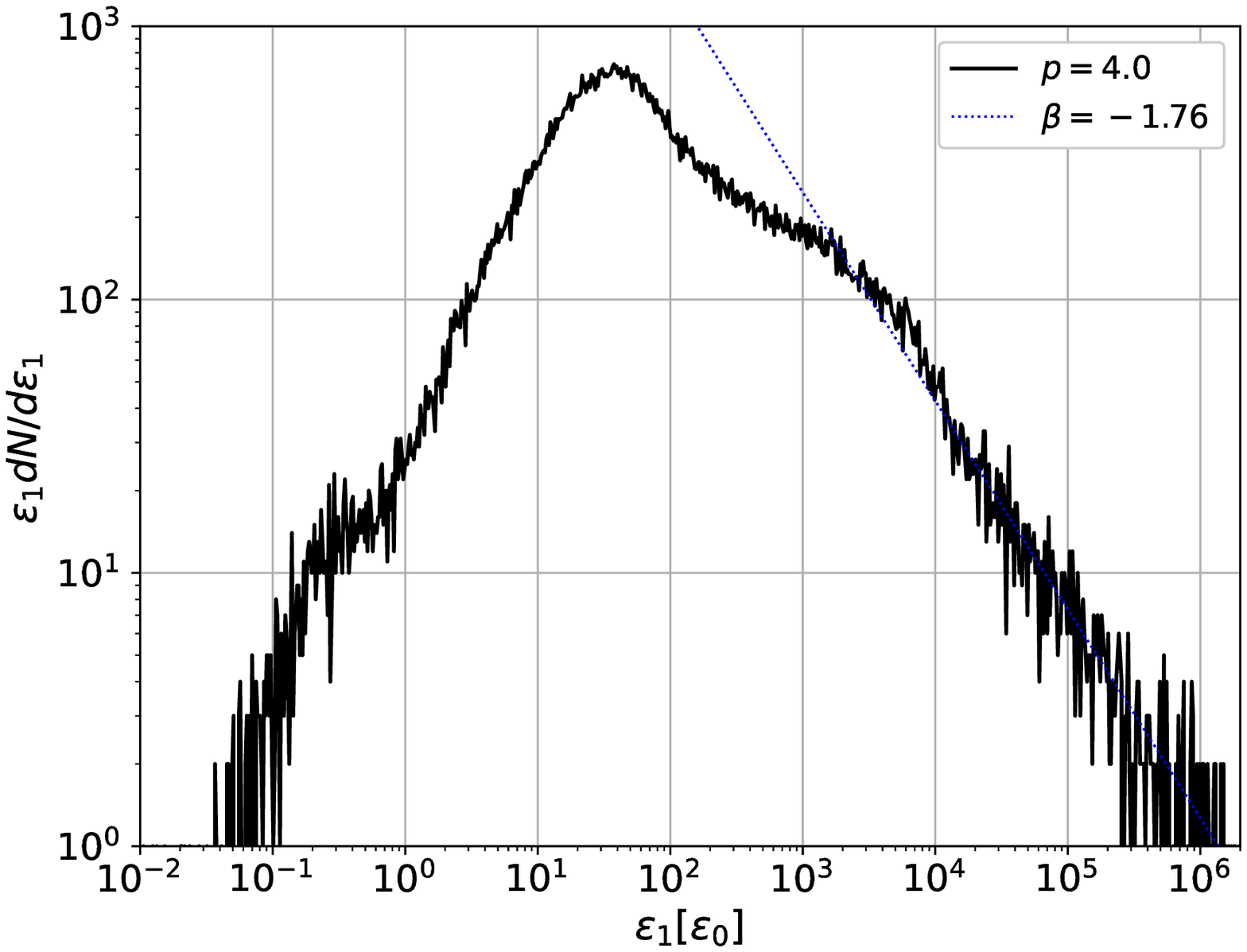} 
            \includegraphics[width=8cm, angle=0]{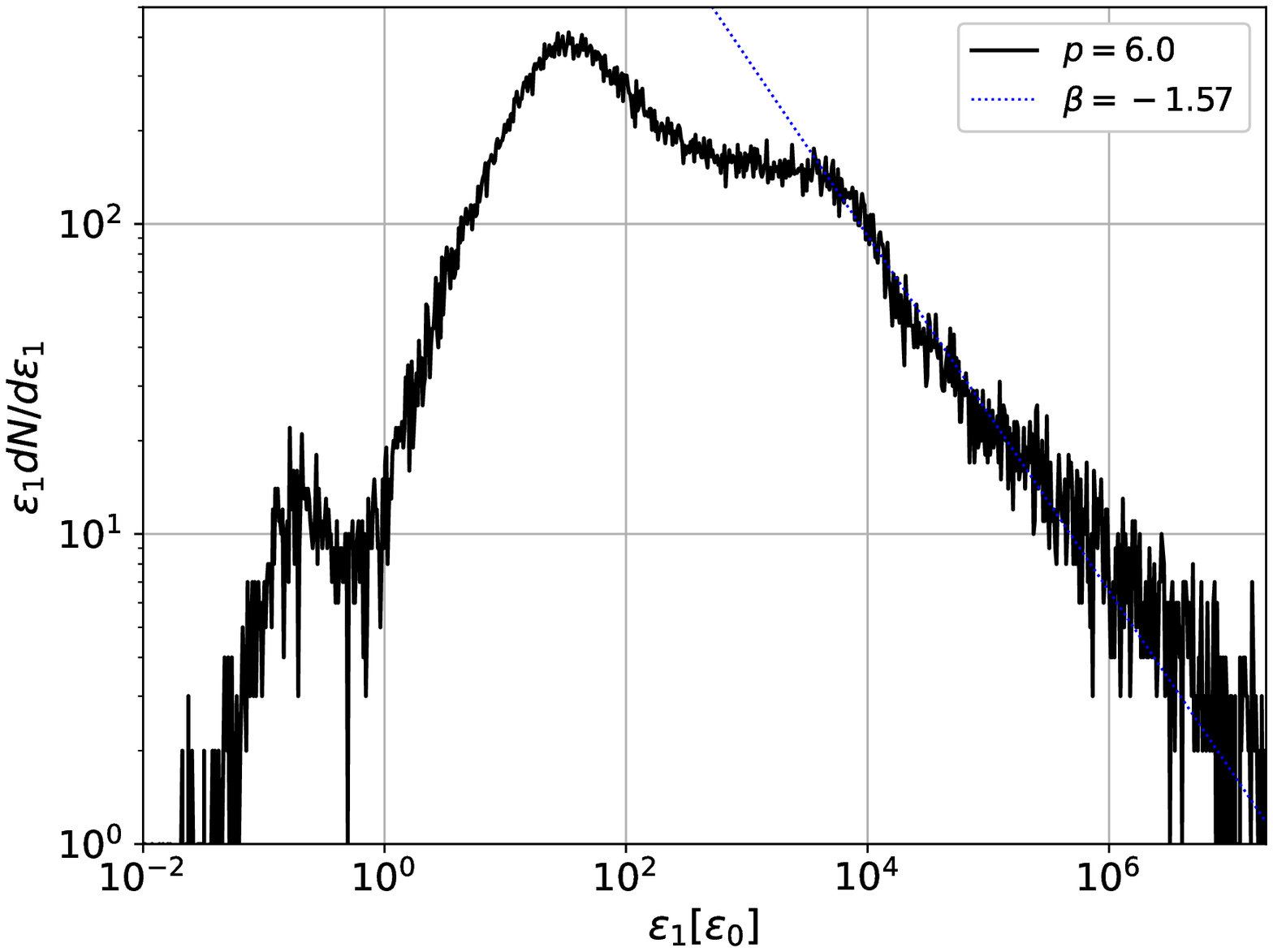}          
             \includegraphics[width=8cm,  angle=0]{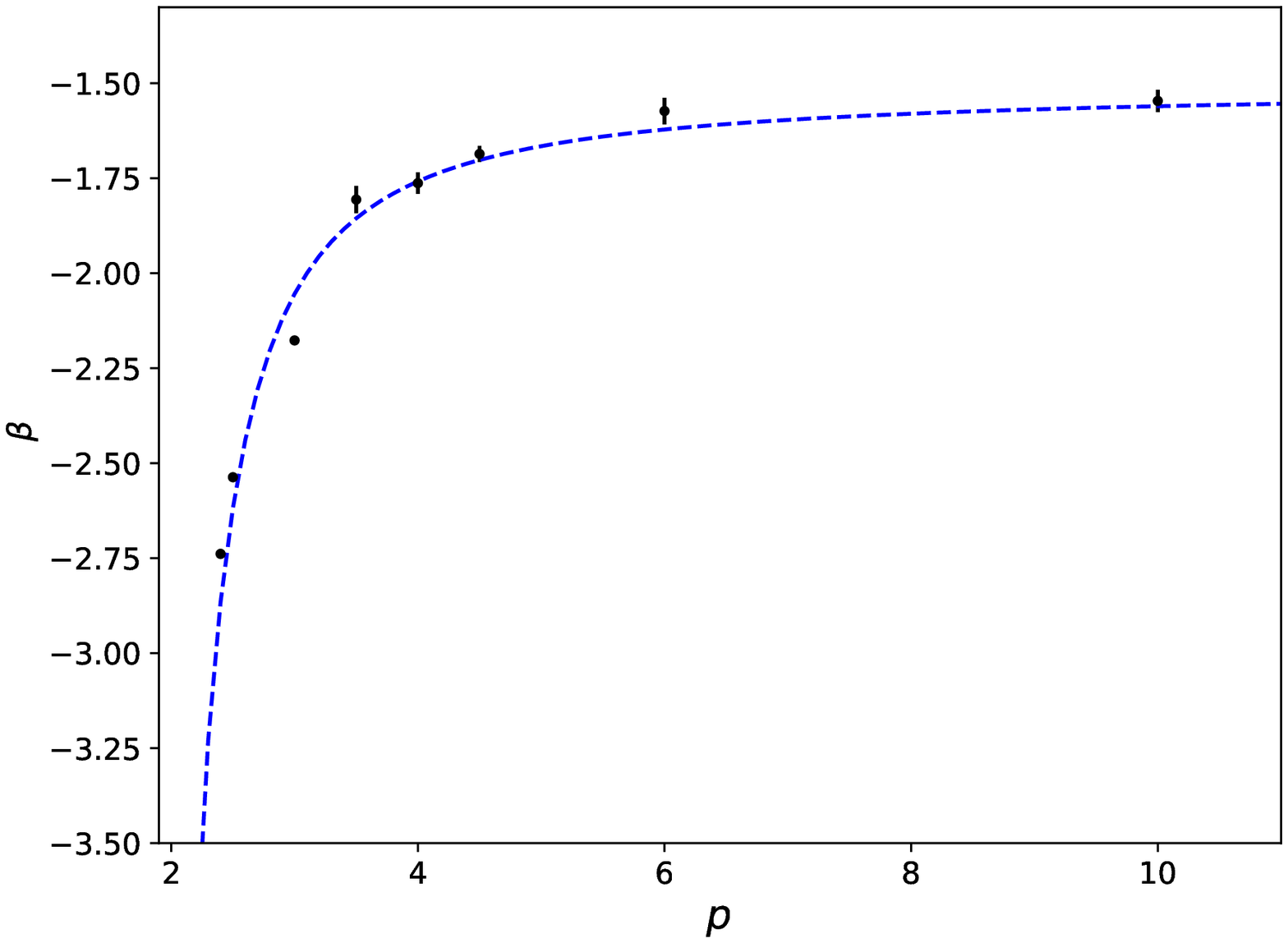}
\caption{Spectra obtained by numerical simulations for $\Gamma=100$, $p=2.5-6$ (solid black) along with fits (dotted blue) for high energy slopes $\beta$. The low energy part of these simulated spectra was modelled by Lundman et. al. \cite{Lundman.etal.2013MNRAS.428.2430L} and we explain the high energy power-law spectrum. The last panel shows the variation of analytically calculated $\beta$ as a function of $p$ (blue dashed). The simulated photon indices are over-plotted by black solid points and are found to be in good agreement with the theoretical curve. The slopes are fitted with power-law indices in the high energy range and the respective error bars in simulated points for $\beta$ are the associated standard deviation. Here $\theta_{\rm j}=0.01$ rad, $\Gamma_{\rm min}=1.2$ and $L=10^{52}$erg$/$s, the monoenergetic seed photons are injected with energy $\varepsilon_{\rm 0}=10^{-6}$ (in the units of the rest mass energy of the electron) in the Monte Carlo Simulation code with identical parameters at radial coordinate $r_0=r_{\rm ph}/d$. Depth $d=20$ ensures that the photons are injected deep inside the flow. The spectra are seen by an on-axis observer at $\theta_{\rm o}=0$ rad (\ie the populations of photons escaped within an angular bin $\theta=0-0.02$ rad). The total number of photons injected is 4 million, 2 million and 2.6 million for $p=2.5, 4.0$ and $6.0$, respectively.}
\label{lab_spectrum}
 \end{center}
\end{figure*}

Assuming a constant Lorentz factor $\Gamma$ along the photon's path, it can be taken outside the integration in equation \ref{eq_tau_2}. This assumption means that while calculating the mean free path at a certain location along a small angle $1/\Gamma$, one can approximate that the Lorentz factor of the outflow encountered by the photon is $\approx \Gamma$. It enables an analytical estimate of $\tau_2$ and is justified because the photon escapes to the inner region ($\theta<\theta_{\rm j}$) only when it is very close to the boundary ($\theta\sim\theta_{\rm j}$) where the Lorentz factor doesn't vary much along the photon's path (Figure \ref{lab_jet_profile}). Noting that $1-v \cos \theta_{\rm s} \approx 1/\Gamma^2$, the optical depth can be written as 
\be 
\tau_2=\frac{2 r_{\rm ph}}{v}\int_0^{s_0}\frac{ds}{r'(s)^2}.
\ee
Here $r'(s)=r+s \cos \theta_{\rm s}$ is the radial coordinate along path $s$.
Direct integration gives
\be 
\tau_2=\frac{2r_{\rm ph}}{v r \cos \theta_{\rm s}}\left[1-\frac{1}{(\theta-\theta_{\rm j})\Gamma \cos \theta_{\rm s}+1}\right].
\label{eq_tau_2_1}
\ee
We use equations \ref{eq_gamma_1}-\ref{eq_tau_2_1} and follow the procedure described above to calculate the observed photon index $\beta$ (Equation \ref{eq_photon_ind}) produced by a GRB jet. The integration over the jetted region $\theta_{\rm j}-\theta_{\rm e}$ and $r_0-r_{\rm ph}$ is performed as there is no significant photon energy gain in the  regions $\theta<\theta_{\rm j}$ and $\theta>\theta_{\rm e}$.

The simulated spectra for $p=2.5,4.0$ and $6.0$ are plotted in the first 3 panels of Figure \ref{lab_spectrum} (see \cite{Pe'er.2008ApJ...682..463P,Lundman.etal.2013MNRAS.428.2430L,Vyas.etal.2021ApJ...908....9V} for the code structure and respective details). The Low energy part of these spectra was theoretically modelled by Lundman et. al. \cite{Lundman.etal.2013MNRAS.428.2430L} while we have modelled the high energy-power law in this letter.
In the last panel of Figure \ref{lab_spectrum}, we compare the photon indices obtained by simulations (dots) with ones calculated analytically (dashed curve) using Equation \ref{eq_photon_ind}. The simulated photon indices are in agreement with the calculated slopes. For the reasons given above, photon's energy loss due to expansion dominates over energy gain by velocity shear for small values of profile index ($p<2$) and there is no high energy power laws below this threshold. 

Due to complexity of the integrals in Equations \ref{eq_int_gain} and \ref{eq_average_P_final}, exact analytic expression of the photon indices in the general case is not possible to obtain. However, one can obtain an analytic expression for the asymptotic behaviour of $\beta$ as $p\rightarrow \infty$ as follows. Assuming $\tau_2\ll 1$ or $P(r,\theta)\sim\tau_2$ analytic integration gives $\bar{g}\propto p^2$ and $\bar{P}\propto(\Gamma_0^{1/p}-1)$ which lead to $\beta\rightarrow -1.5$. This is indeed seen in Figure \ref{lab_spectrum} for both the semi analytic results as well as in the simulated slopes.

Quantitatively, tackling the case of a GRB jet with a structured $\Gamma$ profile as a function of polar angle $\theta$, the Comptonized photons can extend up to several orders of magnitudes. The produced power-law spectrum at high energies is capable of explaining the high energy tails generally observed in the GRB prompt phase. We analyzed that the photon indices of these spectra largely depend upon $p$ which determines the sharpness of the decay of $\Gamma$ with $\theta$. The photon index $\beta$ ranges from $-\infty$ to $-1.5$, which we further confirm by numerical simulations. It is worth mentioning that $\beta$ in GRB prompt phase observations are found to be between $-4$ and $-1.5$ \citep{Preece.etal.2000ApJS..126...19P, Kaneko.etal.2006ApJS..166..298K, pe'er2015AdAst2015E..22P} and this range is in agreement with our results. Inversely, using the observed values of $\beta$, we can directly constrain the jet structure of these bursts.

Although the physical picture of the photon energy gain described here has similarities with the second-order Fermi acceleration of massive particles, important differences must be highlighted. In analytical calculations of the Fermi acceleration process, the expectation value of the gain and the scattering probability are averaged over scattering angles. In the currently discussed mechanism, the energy gain not only depends upon the scattering angle but also on the scattering location. Hence, one needs to average the gain and the scattering probability over the entire scattering region to obtain their expectation values. Furthermore, the escape of photons from the inner jet boundary adds an extra constraint on the escape probability hinting at the anisotropic nature of the energy gain process. This is to be compared to the second-order Fermi acceleration where a locally isotropic scattering is assumed.

We conclude that the commonly observed high energy power-law spectra from various astrophysical sources can have a natural origin due to repeated scattering of photons in relativistic flows with velocity shear. Thus, we provide a novel and viable alternative to the generally considered processes such as synchrotron and inverse thermal Comptonization. These mechanisms need the presence of highly relativistic electrons while in current work the bulk kinetic energy of the jet is transferred to the photons leading to a high energy tail excluding the presence of high energy particles as a priory requirement.

\acknowledgments
\begin{acknowledgments}
AP acknowledges support from the European Union (EU) via ERC consolidator grant $773062$ (O.M.J.). MKV acknowledges the PBC program from the government of Israel as well as part of funding obtained by the above-mentioned ERC grant.
\end{acknowledgments}

\appendix

\bibliography{references}

\end{document}